# Vegetable oil hybrid films cross-linked at the air-water interface: Formation kinetics and physical characterization

Antigoni Theodoratou, [*][†] Laurent Bonnet, [‡] Philippe Dieudonné, [‡] Gladys Massiera, [‡] Pascal Etienne, [‡] Jean-Jacques Robin, [†] Vincent Lapinte, [†] Joël Chopineau, [†] Julian Oberdisse, [‡] Anne Aubert-Pouëssel [†]

[†]*Institut Charles Gerhardt Montpellier (ICGM), UMR5253 CNRS-UM-ENSCM, place Eugène Bataillon, 34090 Montpellier, France.*
[‡]*Laboratoire Charles Coulomb (L2C), UMR5221 CNRS-UM F34095, Montpellier, France*

## Abstract

Vegetable oil based hybrid films were developed thanks to a novel solvent- and heating- free method at the air-water interface using silylated castor oil cross-linked via a sol-gel reaction. To understand the mechanism of the hybrid film formation, the reaction kinetics was studied in detail by using complementary techniques: rheology, thermogravimetric analysis, and infrared spectroscopy. The mechanical properties of the final films were investigated by nano-indentation, whereas their structure was studied using a combination of wide-angle X-ray scattering, electron diffraction, and atomic force microscopy. We found that solid and transparent films form in 24 hours and, by changing the silica precursor to castor oil ratio, their mechanical properties are tunable in the MPa-range by about a factor of twenty. In addition to that, a possible optimization of the cross-linking reaction with different catalysts was explored and finally, cytotoxicity tests were performed on fibroblasts proving the absence of film toxicity. The results of this work pave the way to a straightforward synthesis of castor-oil films with tunable mechanical properties: hybrid films cross-linked at the air-water interface combine an easy and cheap spreading protocol with the features of their thermal history optimized for possible future micro/nano drug loading, thus representing excellent candidates for the replacement of non-environment friendly petroleum-based materials.

## 1. INTRODUCTION

Replacing petroleum-derived materials to avoid the environmental consequences of petroleum extraction has triggered a strong research activity over the past years. The extracted natural oils from plants, consisting mainly of triglycerides, are ideal alternatives to components of petroleum derivatives. For this purpose, since the nineties, vegetable oils have drawn considerable interest by creating biocompatible (harmless to living tissue) cross-linked polymers[1,2,3] and composites,[4] focusing on their synthesis[5] and structural properties.[6,7,8] Bio-based polymeric materials have been applied to several applications, such as coatings,[9,10,11] adhesives[12] and smart polymers.[13] Besides the important positive environmental impact of vegetable-based materials, another interesting property of vegetable oils is their inexpensiveness, their worldwide availability and versatile chemistry.

Moreover, in the biomedical field the potential for vegetable oil-based materials is more pronounced because of their biocompatibility and low levels of toxicity, finding applications as pansements and contact lenses.[4] More specifically, regarding pharmaceutical applications, the recent development of transdermal drug delivery systems target at replacing creams and ointments, avoiding hepatic first-pass metabolism and controlling drug delivery profiles.[12] Common conventional polymeric materials that are used in the biomedical fields up to now include mainly poly(methyl methacrylate) (PMMA), poly(vinylpyrrolidone) (PVP), paraffin gauze and polyurethanes.[14] These polymeric materials shaped as films exhibit good abrasion resistance and physical strength depending on their synthetic process and exhibit biodegradability or/and biocompatibility, but vegetable oils have showed to be excellent to optimize these properties, i.e biodegradability, biocompatibility as well as antibacterial activity.[4,15]

The most common organic-inorganic hybrid materials are fabricated using organosilane coupling agents because they combine unique properties such as brittleness and hardness and they can reinforce vegetable oil-based materials which do not exhibit high rigidity and strength on their own.[16,17,18.]In literature, the properties of hybrid films have been investigated in terms of structure, mechanical and pharmaceutical properties. More specifically, Guo *et* al[12] reported very promising properties of poly(vinyl alcohol) (PVA) hybrid films as transdermal drug delivery systems, exhibiting non-irritation to skin, good performance in drug release and good water vapour permeability properties. Zhang *et* al[6] developed bio-based hybrid films from soybean oil and castor oil using ring opening reactions, the obtained films exhibited high glass transition temperatures, thermal stability and good mechanical properties, such as tensile strength and elongation break. Luca *et* al[11, 19] reported that hardness and tensile properties of hybrid films increase with the amounts of the silica precursors. The most widely used vegetable oils in the area of hybrid materials are linseed oil,[20] sunflower,[21,22] brown soybean oils[23] and castor oil, with the latter one being the only non-edible oil (known as second generation feedstocks), a property that makes its usage very significant because of the tremendous demand of food sources in the new market.[24]

Castor oil derives from *Ricinus communis* plant, belonging to the *Euphorbiaceae* family. It is a renewable, environment-friendly and it is used for the synthesis of eco-friendly resins without modification, as a raw material, and in some cases in its epoxidized functionalization.[11, 19, 25, 26] Concerning castor oil's toxicity levels, rats and mice were exposed to diets containing up to 10 % castor oil showing only minimal indications of toxicity.[27] Also, it is worth to mention that the castor oil seeds contain ricin, a water-soluble toxin, which is denaturised and inactivated during the heating procedure of the oil extraction process. Regarding its molecular chemical composition, it is predominantly derived from ricinoleic acid (C:18 fatty acid~ 87-90%). Its versatile chemistry is attributed to this unsaturated fatty acid and most importantly to the presence of -OH groups (with number average functionality of about 2.7),[28] which make it a key raw-material for chemical functionalization with multiple applications in coatings, adhesives and medical devices.[4] In pharmaceutics, castor oil is used as a pharmaceutical grade inactive ingredient to emulsify and solubilize oils and other water-insoluble drugs.[29] Castor oil is completely soluble in alcohol while other vegetable oils are not, and it can be hydrolysed in presence of enzymatic environment (using a lipase from Pseudomonas sp. f-B-24),[30] while castor oil-based polyurethanes exhibit an excellent hydrolytic stability.[31]

From the above, it may be inferred that castor oil-based hybrid films constitute a promising class of materials targeting at replacing petroleum-based materials. The great majority of the aforementioned works has been performed employing the method of solvent casting and deposition onto solid surfaces to obtain films, using organic solvents and curing at high temperatures.[4, 7, 9-11, 19, 32-35] However, it remains a challenge to fabricate films with a soft chemistry and simple methodology of shaping. In the present study, we address this issue by investigating a novel way to fabricate films based on sol-gel reaction, preventing the use of organic solvents and curing which involves high temperature ($>100°C$) or UV irradiation. This represents a first important step for future applications of vegetable oil hybrid films as scaffolds for model drugs. These films must be solid, transparent, with a controlled thickness, not sticky, hard and water resistant. One of the most innovative features of our approach is represented by its easy implementation: the film formation takes place at the air-water interface, using silylated castor oil. The aqueous surface favours the spreading of the silylated oil because of its impurity-free nature, leading to the formation of smooth and homogeneous films, an easy procedure to scale-up. More specifically, in this work we address three key challenges: (i) exploration of the kinetics of the film gelation at short times (few hours) and film hardening at long times (1.5 month); (ii) characterization of the physical and physico-chemical properties of the films, tuning their hardness via the ratio of the silica precursor to castor oil and (iii) their cytotoxicity towards fibroblast cells.

## 2. EXPERIMENTAL SECTION

**2.1 Materials.** Pharmaceutical grade castor oil (CO; MW = 934 g·mol$^{-1}$) was supplied by Cooper Pharmaceutique. The major fatty acid of castor oil triglyceride is ricinoleic acid and it is sketched in Figure 1. According to the specification sheets, the water content was lower than 0.3%. 3-(Triethoxysilyl)propyl isocyanate (IPTES; MW = 247.4 g·mol$^{-1}$) and dibutyltin dilaurate (DBTDL; MW = 631.6 g·mol$^{-1}$) were supplied by Sigma-Aldrich. Bismuth carboxylate (K-KAT-348) was supplied by King Industries. For the film's degradability studies various organic solvents were used (tetrahydrofurane, dichloromethane, chloroform and acetone) and sodium phosphate buffer (NaH$_2$PO$_4$/K$_2$HPO$_4$) in the presence of lipase enzyme from porcine pancreas, type II 100 units·mg$^{-1}$ (Sigma Aldrich). All the materials were used without further purification. For the cytotoxicity tests NIH 3T3 fibroblast cells (continuous cell line from mouse embryo) were used bought from ATCC.

**2.2 Synthesis of silylated castor oil.** A solvent free synthesis has been employed to functionalize the castor oil with the silica precursor (IPTES) in the presence of a catalyst to create modified castor oil with cross-linkable functions (ICO). A schematic representation of the synthesis is presented in Figure 1. The reactant mixture was heated at 60 °C for 30 min under magnetic stirring. The reaction was performed with a little excess of the isocyanate (NCO) to ensure full reaction of the castor oil's hydroxyl groups catalysed by 0.8% w/w of DBTDL. $^1$H-NMR spectra of castor oil that was used in this work before and after IPTES modification have been reported by Gallon et al.[36] Silylated castor oil has been synthesized using different molar ratio of NCO to hydroxyl groups (OH). X$_r$ parameter refers to the proportion between the molar ratio of isocyanate to hydroxyl groups ($X_r = n_{NCO}/n_{OH}$). The employed ratios included $X_1$, $X_{0.8}$, $X_{0.67}$, $X_{0.5}$ and $X_{0.33}$.

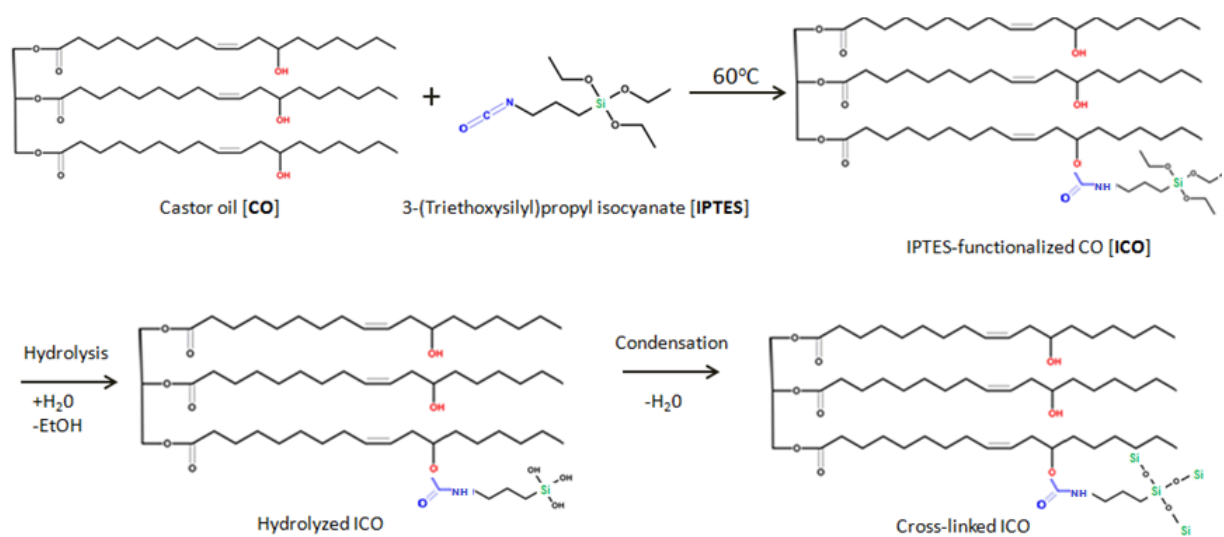

**Figure 1:** Synthesis of IPTES-functionalized castor oil and cross-linking via sol-gel reaction.

**2.3 Protocol for film formation.** The overall procedure of film formation is shown in Figure 2. 1 g of ICO has been deposited on the surface of distilled water (aluminum container-diameter d = 8 cm) at room temperature. Before spreading the ICO, the container was cleaned with ethanol and it was dried at 60°C to avoid impurities at the air-water interface and optimize the spreading of the oil. The system was sealed carefully using insulating tape to ensure 100 % humidity in the above atmosphere, using a humidity monitor that was placed inside the atmosphere to measure its relative humidity. After 2 hours the system reached 100 % humidity and the sol-gel reaction was initialized via 2 steps, hydrolysis and condensation (see Fig.1). The film was left in contact with the water for 3 days and its hardening (cross-linking) continued at room temperature (23°C) for 1.5 month.

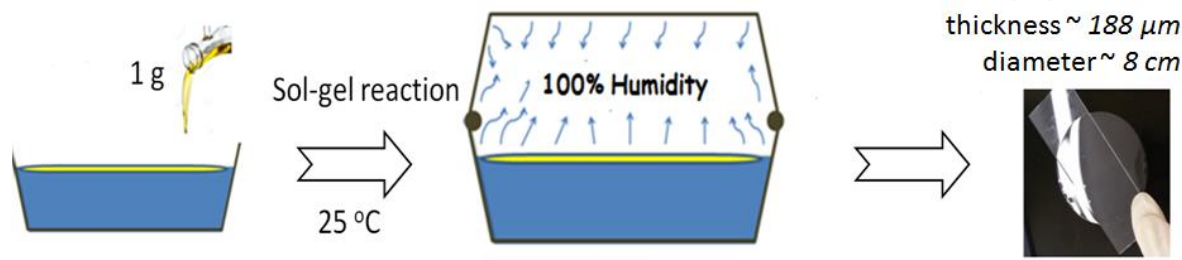

**Figure 2:** Protocol for vegetable oil-based film formation at the air- water interface based on sol-gel reaction.

## 2.4 Characterization techniques.

▪ **Kinetic studies of film formation**

The storage modulus of the film during cross-linking was followed by performing dynamic frequency sweep experiments using a stress-controlled rheometer AR 2000 (TA Instruments) equipped with a steel cone-and-plate geometry (cone diameter = 40 mm). The thermal behaviour of the hybrid films was measured using Thermal Gravimetric Analysis (TGA, STA 6000, Perkin Elmer) from 25 to 600 $^{\circ}$C under air flow and a rate of 5 °C min$^{-1}$ (sample 's mass ~ 10 mg). Attenuated total reflection infrared spectroscopy (FTIR-ATR) spectroscopy (Spectrum 100, Perkin Elmer) was performed in order to follow the kinetics of the cross-linking reaction during the hardening process of hybrid films. Also, the loss of mass of a whole film of 0.86 g during its hardening at room temperature was measured for 52 days.

▪ **Characterization of final films**

Wide angle (WAXS) and small angle (SAXS) X-ray scattering were performed with an 'in-house' setup (at room temperature). A high brightness low power X- ray tube, coupled with aspheric multilayer optic (GeniX3D from Xenocs) was used delivering an ultralow divergent beam (0.5 mrad). Scatterless slits were used to give a clean 0.8 mm beam diameter of 35 Mphotons s$^{-1}$ at the sample. The experiments were done in transmission configuration and the scattered intensity was measured by a Schneider 2D image plate detector prototype, at a distance of 1.9 m from the sample for SAXS configuration and 0.2 m from the sample for WAXS configuration ($\lambda$=1.54 Å). After the collection of the raw data, all intensities were corrected by transmission and the empty cell contribution was subtracted. Scanning electron microscopy (SEM) images of the films were obtained using Hitachi S4800 High resolution microscope and the composition and homogeneity of the surfaces of the films were characterized using Energy Dispersive X-ray (EDX), Hitachi S4500 I, with platinum used as conductor material. The films were analyzed using atomic force microscopy (AFM) (Nanoman V/ Bruker Instruments, Nanoscope 5 Controller) in tapping mode under air at room temperature. The tip had a conical shape and it was mounted on a cantilever (type FM Pointprobe Nanosensors), with resonance frequency of 68 kHz and spring constant of 2.5 Nm[-1]. All images were flattened and analysed using Gwyddion Software (www.gwyddion.net).

[29]Si solid state NMR experiments were performed on a Varian VNMRS 400 MHz (9.4 T) NMR spectrometer, using a 7.5 mm Varian T3 HX MAS probe and spinning at 5 kHz. Single pulse MAS (magic angle spinning) experiments were carried out, using a 2μs 30° pulse. A recycle delay of 60 s was used. For Cross Polarization MAS (CP-MAS) experiments, a 1.5 ms contact time was used with a recycle delay of 5 s. The nano-indentation measurements have been performed at room temperature with an Anton Paar ultra-nano-indentor using a Berkovich diamond tip. The maximal force of indentation was 300 μN with an indentation speed of 150 μN min[-1]. The spacing between each indent was fixed to 30 μm to prevent any interaction. Contact angle measurements were performed with a DataPhysics OCA – Series device using as solvents distilled water and diiodomethane. The droplet shape onto the hybrid films was captured with a CCDs camera to extract the contact angles that were the average of at least 10 measurements. The surface energy was calculated according to Owens, Wendt, Rabel and Kaelbles method.[37,38]

The degree of swelling (Q) of the films in contact with organic solvents was evaluated by measuring the weight of the films before and after their exposition to various solvents according to the following relation:

$$Q = \left(\frac{m_f - m_i}{m_i}\right) * 100$$

where $m_i$ is the initial mass and $m_f$ the final film mass. The films were merged into the organic solvents for 24 h at room temperature. After removing the excess of solvent, they were carefully cleaned with tissue paper and weighted. In order to obtain an average value, three tests were performed for each individual solvent. For the biodegradability tests, according to Yamamoto *et* al,[30] the films were immersed into a 0.1 M sodium phosphate buffer solution (pH=7) mixed with lipase enzyme at a concentration of 13 μg mL at 37 ºC . After 46 days the films were removed from the buffer and they were left to dry for 24h at 37°C and weighted. Three biodegradability tests were performed in order to obtain an average value.

- **Cytotoxicity tests**

For the cytotoxicity evaluation of the hybrid films, NIH 3T3 fibroblasts in cell culture were chosen to be studied with the method of CellTiter 96[®] AQ cell proliferation assay (Promega) made by a tetrazolium compound (3-(4,5-dimethylthiazol-2-yl)-5-(3-carboxymethoxyphenyl)-2-(4-sulfophenyl)-2H-tetrazolium, inner salt; MTS) and an electron coupling reagent (phenazine methosulfate; PMS). Films were mixed at different concentrations (0.1, 1 and 10 mg·ml$^{-1}$) with a culture medium DMEM (Dulbecco's Modified Eagle Medium) and they incubated for 80 days at 37°C. Cytotoxicity measurements were performed at different incubation times (after 2 days, 7 days, 14 days, 28 days and 80 days). The cells were placed into a 96-well culture dish plate (5000 cells per well), left to adhere for 24 h at 37 °C in presence of $CO_2$ (5 %). Then, 100 μL of DMEM (that was previously mixed with the film) were added on the cells. After 48 h of incubation, 20 μL of MTS-PMS mixture was added to the cells and after 4h the percentage of living cells was evaluated. The percentage of the living cells was quantified using a microplate reader (Multiskan Go, Thermo Fisher Scientific) at 490 nm. The same protocol was used for the cytotoxicity measurements of pure catalysts (DBTDL and Bismuth carboxylate) at different concentrations and incubation times.

# 3. RESULTS AND DISCUSSION

## 3.1 Kinetics of castor-oil film formation

Functionalized castor oil (ICO) was deposited on the water surface to induce the sol-gel reaction and gelation of the oil within few hours (Fig. 2). The final films were solid and transparent with an average diameter of 8 cm and thickness of $188 \pm 10$ µm. In order to characterize the evolution of the cross-linking we investigated the rheological, thermal and chemical properties of the film as a function of time. Two different characteristic times have been introduced: (1) the gelation time $t_a$ measured by rheology and characterizing the short-time kinetics; (2) the exponential decay time $t_c$ of hardening, measured with FTIR-ATR and weight loss (long-time kinetics).

**Gelation at short-times.** Shear rheology was performed in order to investigate the cross-linking kinetics of the functionalized ICO. Eleven films were prepared at the same time, and subsequently the gel fraction of each system was collected at different times. Dynamic frequency sweep measurements were performed on the different gel-like films at strain amplitude of $\gamma_0 = 1\%$, in the linear viscoelastic regime (see supporting information -Figure S2). The results of the dynamic frequency sweep tests are shown in Figure 3A, where the elastic modulus (G') is plotted as a function of time at fixed angular frequency of $\omega = 10$ rad s$^{-1}$. It is important to mention here that the plateau observed for t > 24 h do not correspond to the final state of the films because they are in a gel-like state (without complete hardening). Thus, due to the very slow evolution of the moduli, bulk rheology cannot be used to follow the evolution of the cross-linking kinetics after 24 hours. The final mechanical properties of the final films will be described later by means of nano-indentation.

The effect of the temperature on the film gelation was studied using the same method. Since at room temperature the gelation of the film occurs after 24 hours, films were synthesized at controlled temperature for 1 day. The gel-like films which were collected after 24 hours were measured using shear rheology. Dynamic frequency sweep tests were performed at $\gamma_0=1\%$ from which we extracted at 10 rad/s the storage modulus G' as a function of the synthesis temperature (Figure 3B). One order of magnitude increase has been observed for G' from $10^4$ Pa to $10^5$ Pa from 23 °C to 80 °C. Thus, temperature favours the sol-gel reaction, accelerating the degree of cross-linking. Also, the same experiments were performed at T=5 °C. In this case the cross-linking kinetics dramatically slowed down and no solidification occurred after 24 hours.

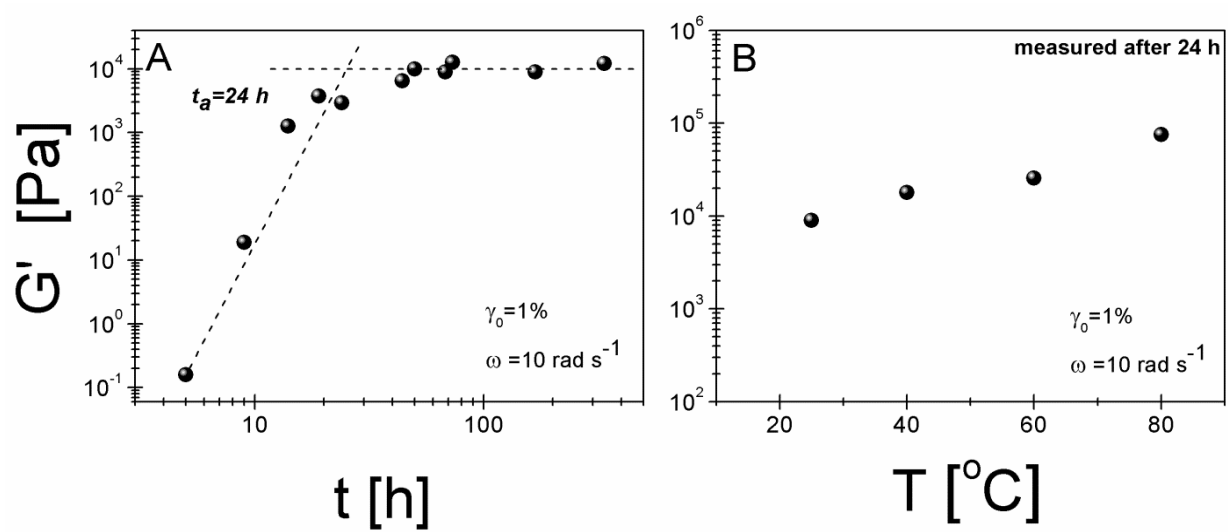

**Figure 3**: Shear rheometry during film gelation. Left panel: Short-time kinetics of film formation represented by the storage modulus (G') as a function of time. Right panel: Temperature effect on film formation. G' is plotted at angular frequency of $\omega=10$ rads$^{-1}$ and strain amplitude of $\gamma_0=1\%$. The storage modulus is extracted from dynamic frequency sweep data shown at Fig. S2. The error bars

are at the size of the symbols.

**Hardening at long-times.** The cross-linking of the films at room temperature during their hardening is studied by means of thermogravimetric analysis. The film has been removed from the water surface after 3 days and TGA measurements were performed every day from different pieces of the same film for 12 days. The results are reported in Figure 4A as percentage of weight loss versus temperature for different times. The film's weight loss up to 600 $^{\circ}$C shows the presence of 12 % inorganic material, which confirms the formation of hybrid material and it is in agreement with the theoretically calculated value (see supporting information Fig. S3a). This result is in agreement with polyurethane/siloxane castor oil films previously reported by Meera *et* al.[39] To explain further the role of the cross-linking on the film's hardening, a fixed temperature (270 $^{\circ}$C) and weight loss (10 %) have been selected to plot the corresponding weight loss and temperature, respectively, as a function of the time. These values have been chosen carefully to be within the regime where no chain scission of oil occurs (first step-initiation of pyrolysis).

In order to characterize the ageing kinetics via TGA we fix a threshold of 10% weight loss and monitor the position of the cross-linking point of all TGA curves with such threshold. In this way we can plot the temperature that corresponds to 10% weight loss as a function of time. As one can see in Figure 4B, the temperature is found to increase first linearly with time, and then stabilizes after 7 days. Superimposing in Figure 4b the weight loss at fixed temperature versus time, it is seen again that after 7 days a plateau is reached. Another fixed weight loss at 0.01 % gave the same time of 7 days (data available at Figure S3b). The above results obtained by thermogravimetric analysis did not shed light into the quantitative degree of the cross-linking, but they provide a meaningful information regarding the time at which a hard (handleable - optimally cross-linked network) film is formed.

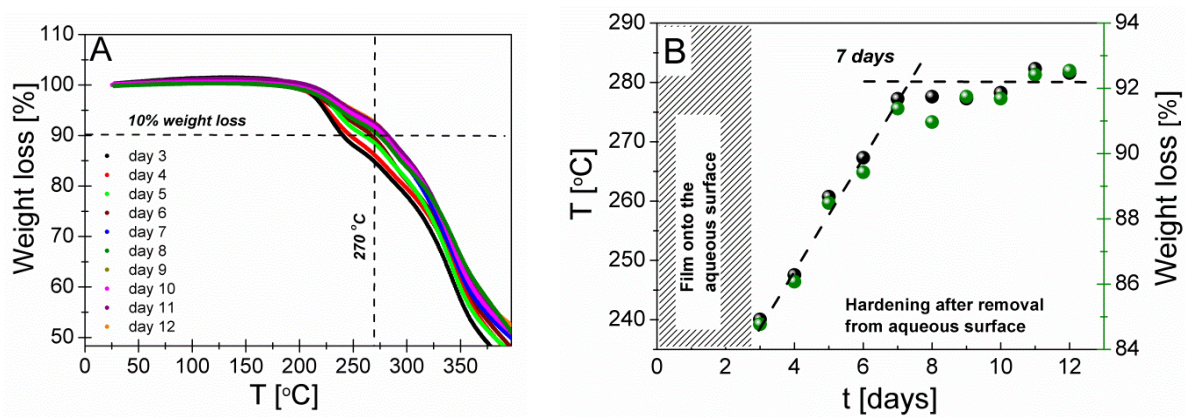

**Figure 4:** Long time kinetics of film formation followed by TGA during hardening procedure. a) Weight loss as a function of temperature (larger scale data up to 600°C are shown at *SI*). b) Temperature at constant weight loss (10%) versus time (black symbols) and weight loss at constant temperature (270°C) versus time (green symbols). The error bars are included at the size of the symbols.

A crucial point for the investigation of the cross-linking kinetics was the estimation of the end of the sol-gel process. For this purpose, the long time kinetics was captured, up to 52 days, using infrared spectroscopy (FTIR-ATR) and the weight loss of the film at room temperature after its removal from the water surface. Figure 5A depicts the collection of transmittance spectra versus wavelength at different times (0<t<52 days) at wavelengths between 975-1150 cm⁻¹. The complete spectra 500 <λ< 4.000 cm⁻¹ are shown in supporting information (Fig. S4). A peak at λ= 1076 cm⁻¹ characterizes the spectra and is typical of ethanol formed during the hydrolysis reaction (Figure 1).[40] Looking at the spectra of Figure 5A, an obvious decrease of peak's area at 1076 cm⁻¹ is observed which stabilizes with time. For better visualization of the time dependence of this peak, the integrated area of the peak is plotted as a function of time (Figure 5B). The evolution of the peak corresponding to Si-O-Si and Si-OH bond as a function of time (see supporting information Figs S4-B and S4-C) is the signature of the ongoing cross-linking reaction. ²⁹Si solid state NMR spectra of the final film (after 52 days) are presented in supporting information (Fig. S1), which confirm the formation of siloxane network.

To verify the previous observation on the long-term kinetics, the mass of the films was measured and the results are reported in Figure 5B, displaying a very good agreement with the FTIR-ATR results. In summary, a prolonged sol-gel reaction favours the cross-linking accompanied by ethanol production (that escapes) and could be measured by weighting the films. The results presented in Fig. 5B have been fitted using an exponential fitting function $M(t) = M_0 e^{-t/t_c}$ for the temporal evolution of the mass, where the decaying time $t_c$=13.7 days represents the time scale characterizing the sol-gel reaction within the film. We have estimated stoichiometrically the ethanol product during the hydrolysis step of the cross-linking reaction which gives a value equal to the 22.8 % of the initial mass of silylated oil (8.1 mol of produced ethanol for 1 mol of silylated oil, with molecular mass of 46 and 1634 g·mol$^{-1}$, respectively). As it was described above, the ethanol loss could be monitored by measuring the weight of the film in function of time. More specifically, the initial mass of the silylated oil that we deposited on the water surface was 0.976 g and after its complete hardening (52 days) the mass of the film decreased down to 0.764 g, corresponding to 21.7 % of weight loss with respect to the initial mass of oil (0.976 g). This is a result in good agreement with that obtained from stoichiometric calculations.

Consequently, from rheology we have obtained the typical time of the oil gelation $t_a$=24 hours; from TGA we observed that after 7 days a hard (and handleable) film is obtained, and finally $t_c$=13.7 days from FTIR-ATR and weight loss studies referring to the decaying time of the cross-linking reaction. At this point it is worth to mention that a film removed from the water surface after 3 days and cured at 120 °C overnight has the same macroscopic properties of a film left to cross-link at room temperature for t>$t_c$. Another way to accelerate the cross-linking reaction is the replacement of DBTDL catalyst with a sulfonic acid which is one of the most important future perspectives of our work.

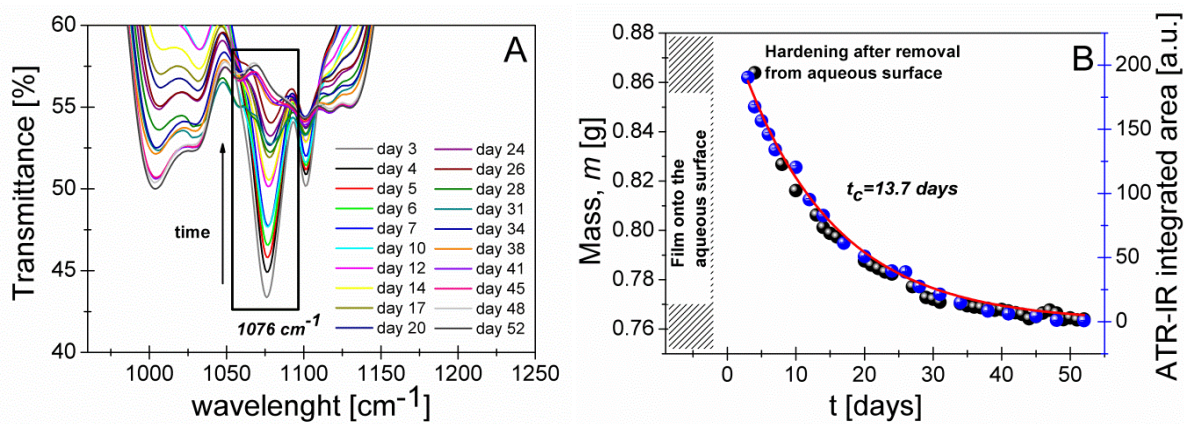

**Figure 5:** Long time kinetics of film formation during its hardening followed by (FTIR-ATR) and weight loss. a) Transmittance versus wavelength, b) Integrated area of the peak 1076 cm$^{-1}$ (blue symbols), film mass (black symbols) as a function of time and its exponential fit (red line). The error bars are given by the size of the symbols.

## 3.2 Characterization of castor oil hybrid films

### 3.2.1 Structural properties of CO hybrid films

The spatial organization (i.e. intermolecular) of the raw castor oil, its functionalized oil ICO, and the cross-linked film have been characterized by WAXS. In Figure 6, the amorphous structure of raw and functionalized castor oil, as well as cross-linked films are shown. Note that the intensity is not normalized, but the curves have been superimposed at large angles. A strong peak is found for the pure oil at q = 1.4 Å$^{-1}$ (corresponding to 2θ=20º), which is found to decrease with functionalization and cross-linking. The characteristic peaks of castor oil polyurethanes [32,41] at q=1.4 Å$^{-1}$ correspond to a molecular distance of d = 4.5 Å, evaluated according to the following Bragg's equation:

$$d = \frac{2\pi}{q} \ (\text{Å})$$

where $q\left(\text{Å}^{-1}\right) = \frac{4\pi \, \sin\theta}{\lambda}$ is the wavevector of the incident beam.

This peak has been related to the presence of short range local order of the triglycerides in coexistence with a disordered amorphous phase of the polyurethanes. Furthermore, we may also point out that pure castor oil shows the highest number density of ordered domains, while its functionalization introduces more disorder that stays frozen once full cross-linking occurs. We should note that small-angle X-ray scattering (SAXS) experiments and small angle neutron scattering (SANS) have also been attempted (see supporting information, Fig. S5). However, no long range ordering was found in the low-q range (nanometer and tens of nanometer scale). Thus, the sol-gel cross-linking do not create crystalline structure of the hybrid films.

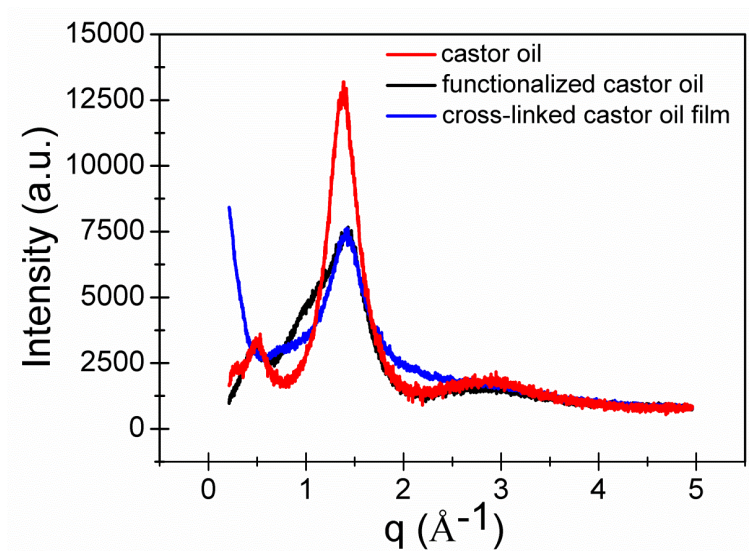

**Figure 6**: Structural characterization of the castor oil, IPTES-functionalized oil and the cross-linked film measured by wide angle X-ray scattering.

AFM measurements have been performed both on the surface of the films, and on their internal structure (bulk) by cutting manually the films. The 3D images are presented in Figure 7-A,B for the bulk structure and the surface, respectively. It was found that the films are homogeneous inside with a roughness of 8 nm, and carry a number of defects on the surface with an average roughness of 4 nm. The films present more defects on the air-film (top) interface comparing with the film-water (down) interface. We exclude that the defects are due to pre-existing bubbles within the oil, since the functionalized oil was subjected to degassing by sonication. Hence, these defects could originate from the ethanol produced during the cross-linking reaction.

The homogeneity of the interior part of the films has been visualized by means of scanning electron microscopy and it is presented in Figure 7C. The bulk interior part of the film seems smooth without a visible porosity on the micrometric scale. Furthermore, the thickness of the films has been found to be $188 \pm 10$ μm. EDX analysis was conducted in parallel with SEM measurements in order to confirm that a hybrid material has been formed. The average experimental values for the percentage of the atomic masses is presented in the inset of Figure 7D, showing a homogeneous film without separated phases of organic and inorganic parts.

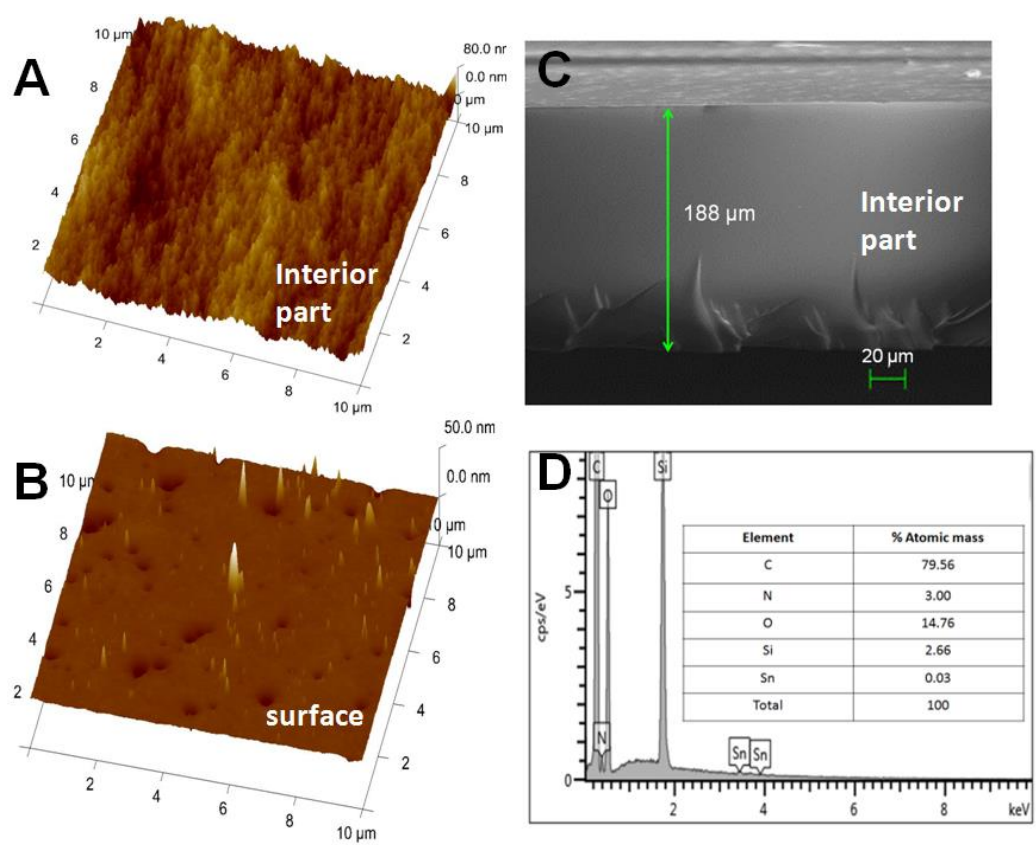

**Figure 7:** Structural characterization of CO films: Atomic force microscopy images of the interior part of the film (A) and its surface (B). Scanning electron microscopy image of the interior part of the film (C) and EDX analysis of the film surface.

### 3.2.2 Mechanical properties of CO hybrid films

Nano-indentation experiments have been performed to study the mechanical properties of the hybrid films such as Young's modulus and hardness. The films that were used had a thickness of $188 \pm 10$ μm and a maximum penetration depth of 2 μm was used. The obtained data include the loading (P) as a function of the penetration depth (h) curve until a predefined maximum force of 300 μN and the unloading curve. The hardness (H) is calculated at the maximum force according to the following relation:

$$H = \frac{F}{A_P}$$

where F is the maximum force and $A_P$ is the surface contact area between the indenter and the sample. The Young's modulus (E) of the sample can be obtained using the slope (S) at the beginning of the unload curve according to the following relations:

$$E^* = \frac{\sqrt{\pi}}{2\beta} \frac{S}{\sqrt{A_P}} \quad \text{and} \quad \frac{1}{E^*} = \frac{1 - \nu^2}{E} + \frac{1 - \nu_i^2}{E_i}$$

where β is a geometrical constant depending of the shape indenter (1.034 in our case), $E_i$ and $\nu_i$ are the indenter Young's modulus and the Poisson coefficient respectively, ν is the Poisson coefficient of the sample. Poisson coefficient is fixed at a value of 0.4 for all measurements as a compromise between polymers and elastomers corresponding to the mean behaviour of our films.

Two series of hybrid films have been synthesized and characterized by means of nano-indentation. First, films have been made with various IPTES/castor oil ratio ($X_r$). In Figure 8a, the load-unload curves as a function of depth for different $X_r$ ratios and their corresponding Young's modulus and hardness are presented in Figure 8b. In Figure 8a a horizontal shift is observed for smaller $X_r$, meaning a lower cross-linking ratio and softer films. As a consequence, the Young's modulus for $X_1$ reaches 85 MPa and it decreases with decreasing $X_r$ ratio. More specifically, the Young moduli are: $E_{0.8}$=63 MPa, $E_{0.67}$= 27 MPa, $E_{0.5}$=11 MPa and $E_{0.33}$=4.3 MPa. These values are comparable with the Young modulus observed by de la Rosa-Fox *et* al[42] for hybrid organic/inorganic silica aerogels. The hardness decreased from 22 ($H_1$) down to 1.3 MPa ($H_{0.33}$) accordingly. As a result, the mechanical properties of the hybrid films can be tuned by modification of the $X_r$ ratio with Young modulus between $4.3 < E(MPa) < 85$.

Second, as already mentioned in section 3.1, hybrid films were synthesized at different temperature 23, 40, 60 and $80^\circ$C, using in all cases $X_r$=1, and their storage modulus was measured after 24 h of their deposition on the water surface. The final mechanical properties of these films were characterized with nano-indentation after $t_c$>13.7 days. In Figure 8c, the load-unload curves as a function of depth for different temperatures are presented and their corresponding Young's modulus and hardness in Figure 8d. The results do not indicate any particular dependence of the Young modulus and hardness on temperature at which the films have been synthesized. It is interesting to note that the Young's modulus is $10^4$ orders of magnitude larger than the storage modulus observed from rheology for the gel-like samples. This is not surprising given that the films are in their early state of cross-linking after t=24 h while their cross-linking decay time was found to be after 13.7 days. Consequently, the final mechanical properties of the hybrid films are not affected by their formation at different temperatures.

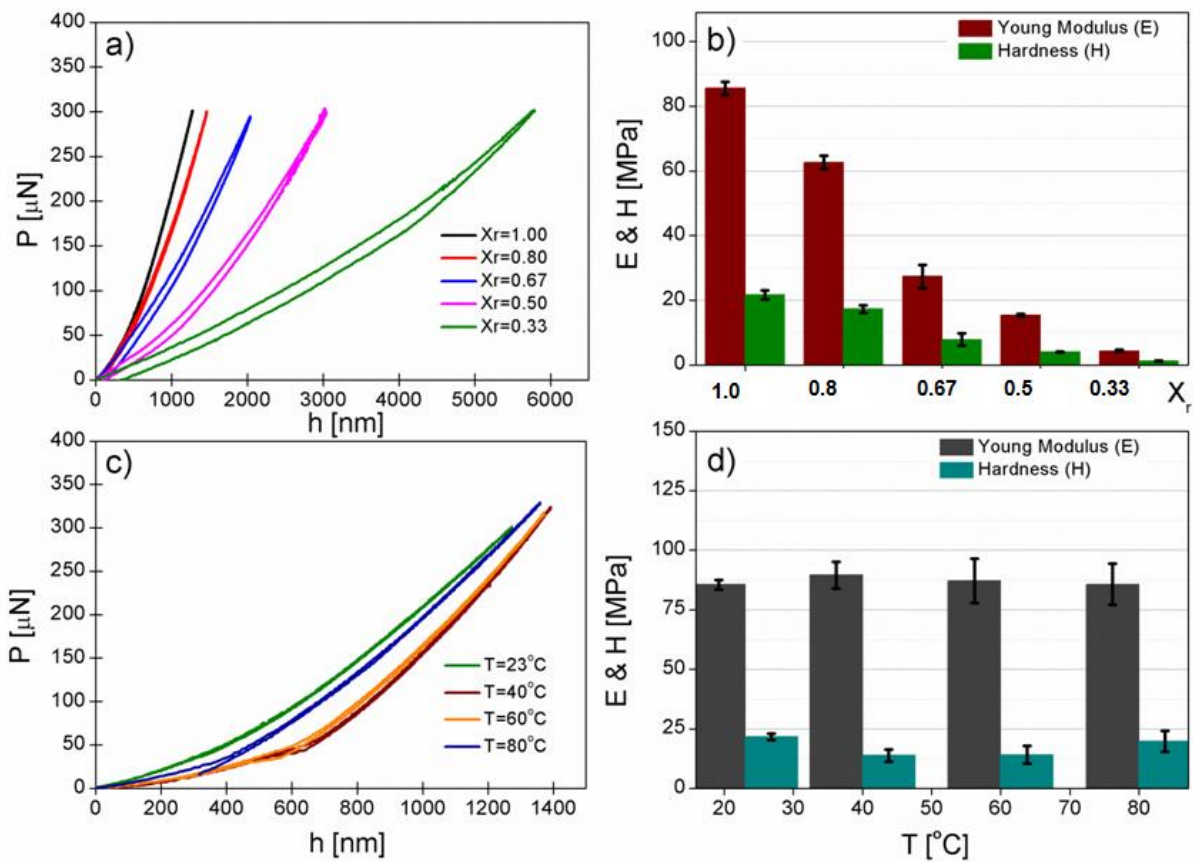

**Figure 8:** Mechanical properties of final hybrid films by nano-indentation: (a) Load curves as a function of depth with different $X_r$ ratios and (b) resulting Young's modulus and hardness. c) Load curves as a function of depth for films synthesized at different temperatures and d) their resulting Young's modulus and hardness. The maximum load in all cases was 300µN.

### 3.2.3 Physico-chemical behaviour of CO hybrid films

**Contact angle measurements.** The solvophobicity of the hybrid films was studied by contact angle measurements. Two series of measurements were performed using distilled water (hydrophilic solvent) and diiodomethane (hydrophobic solvent). Diiodomethane was chosen because of its low volatility and low chemical reactivity. The liquids were deposited on the film substrates and formed spherical drops with average contact angles of 85.5 ± 3.4º and 55.2±3.1º for water and diiodomethane, respectively (Fig. 9). The contact angle found for the water drop is in agreement with literature,[4] for castor oil films made using the method of solvent casting. Thus, the silica of the hybrid material did not modify the global hydrophobicity of the film. In the case of the water drop $\theta$ =85.5 °<90° pointing out the hydrophilic character of the film surface. We found the interfacial energy of the film/air interface was $\Gamma_{f/a}$= 32.1 mNm$^{-1}$.

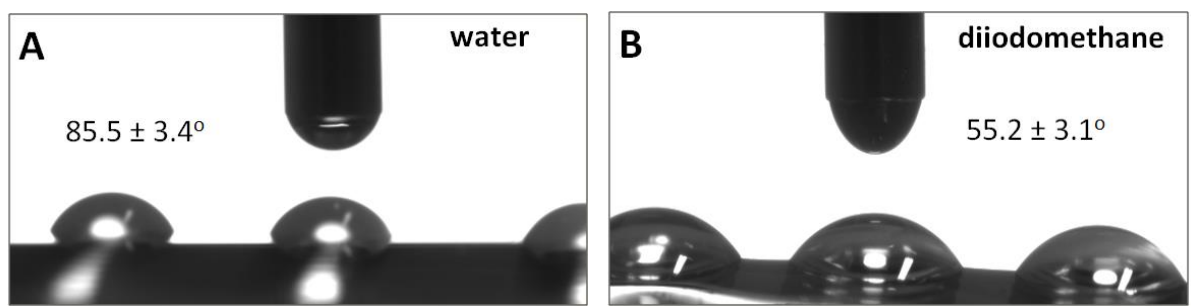

**Figure 9:** Contact angle images for drops of water (A) and diiodomethane (B) on CO films.

**Resistance of CO hybrid films.** The resistance of the films was assessed by using different organic solvents (tetrahydrofurane, dichloromethane, chloroform and acetone) to obtain information regarding the quality of the cross-linking of the hybrid network towards solvents. It was found that the films (after $t_c$>13.7 days) do not degrade significantly when immerged in any of the organic solvents mentioned above. Furthermore, the degree of swelling was estimated to be 5±1%, indicating the high cross-link density of the films. Finally, biodegradability tests using a phosphate buffer with enzyme (lipase) did not show any material loss after an incubation time of 46 days. Consequently, we conclude that the hybrid films made by castor oil are particularly resistant in the presence of organic solvents or water and were not enzymatically degraded in our experimental conditions.

### 3.3 Cytotoxicity with fibroblasts and catalyst replacement

The cytotoxicity of the hybrid films was evaluated via standard viability tests using fibroblast cells. The cytotoxicity of the films was studied at different incubation times (from 2 days to 80) and different film concentrations of 0.1, 1 and 10 mg·ml$^{-1}$. The results shown in Figure 10A are expressed by the percentage of cell viability versus incubation time for the different concentrations. The control sample (0 mg·ml$^{-1}$) is used as a reference system corresponding to cell viability of 100%. A good cell viability has been observed for concentrations of 0.1 and 1 mg·ml$^{-1}$ for all the examined period of time, for 2 months. It means that hybrid films do not release toxic elements possibly affecting the viability of the fibroblasts. At the higher concentrations (10 mg·ml$^{-1}$), when the films incubate with the cells for more than 7 days, a decrease of the cytocompatibility is observed from 67 to 45 % after 2 months. It is strongly believed that the traces of DBTDL catalyst can be released with time and possibly increase the observed toxicity. Looking towards this direction, several attempts were made to replace the tin catalyst with a more environment-friendly catalyst. Bismuth carboxylate was found to be a good candidate to synthesize castor oil hybrid films. In Fig. 10B we present a comparison between the two catalysts at the same concentrations (0.01, 0.1 and 1 mg·ml$^{-1}$) incubated for 1 and 7 days. It is shown that the cytotoxicity levels are dramatically decreased by replacing DBTDL with Bismuth carboxylate. The perspectives of future work with bismuth carboxylate include the study of the kinetics of cross-linking as well as optimization of the film formation procedure.

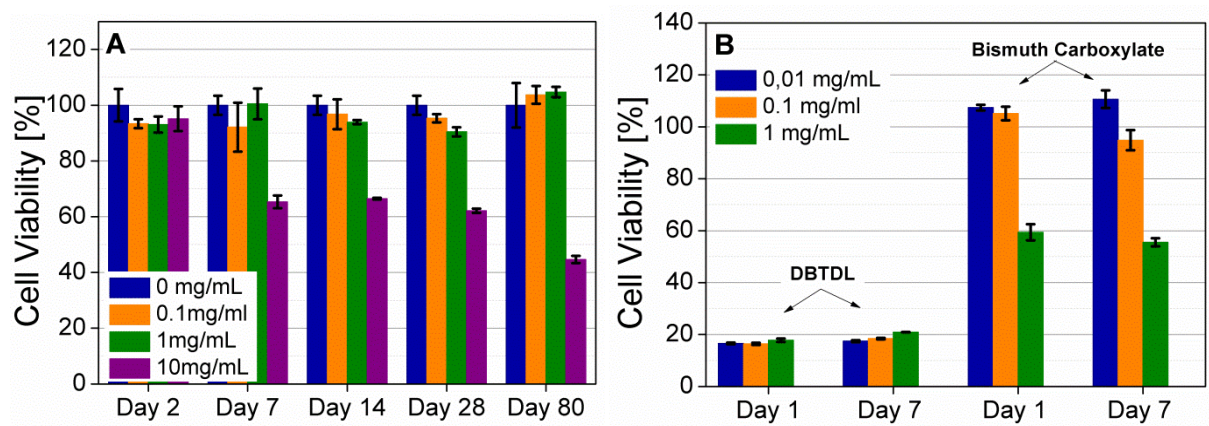

**Figure 10:** Cytotoxicity of a) the hybrid films (using DBTDL) and b) pure catalysts (DBTDL and Bismuth Carboxylate) using fibroblasts expressed by a percentage of cell viability as a function of incubation time at different concentrations.

## 4. CONCLUSIONS

Hybrid castor oil films have been synthesized at the air-water interface in soft conditions i.e. without the use of organic solvents and curing at high temperature. The kinetics of the cross-linking of the film has been studied, as well as the characterization of the final films in terms of structure, mechanical properties, physico-chemical properties and cytocompatibility. The results of kinetic studies pointed out that long times are needed for the completion of the cross-linking reaction: Different time scales were investigated by: 1) rheology ($t_a$=24 hours needed for gelation); 2) infrared spectroscopy and weight loss (exponential decay time $t_c$=13.7 days). The kinetics of film formation was studied at room temperature in order to avoid high-temperature curing for future applications of the films as scaffolds for model drugs. From X-ray scattering the structure of the film was probed at wide angles, where a short range local ordering of the triglycerides in coexistence with a disorder amorphous phase was shown, while no long range order was found in the low-q range at small angles. The homogeneity of the films was confirmed by AFM and SEM studies showing no porosity in the cross-linked matrix. TGA and $^{29}$Si solid state NMR confirmed the existence of hybrid inorganic/organic material showing 12 % of inorganic part and the presence of siloxane bonds. The mechanical properties measured by nano-indentation allowed to estimate the Young modulus and the hardness of the films at different percentages of silica precursor and different temperatures at which the films are synthesized. In brief, the results demonstrated that Young's modulus and hardness can be controlled by increasing the silica precursor, while increasing the temperature at which the films have been synthesized, the final mechanical properties do not change. The degree of hydrophilicity(-phobicity) of the hybrid films is of great importance for applications, for this reason contact angle measurements were performed and the results revealed that films are hydrophilic, with contact angles in agreement with those of castor oil films previously reported by Diez-Pascual *et* al and Meera *et* al.[4,39] Finally, the pharmaceutical potential application of the castor-oil films has been confirmed by the high fraction of living fibroblast cells obtained by systematic cytocompatibility tests, and the successful replacement of tin catalyst with a non-toxic one.

- **AUTHOR INFORMATION**
  *E-mail: antigoni.theodoratou@umontpellier.fr

- **ACKNOWLEDGEMENTS**

We thank Michel Ramonda for AFM, Bertrand Rebiere for SEM/EDX, Dafne Musino for SANS experiments and Philippe Gaveau for $^{29}$Si solid state NMR experiments. The authors are grateful for the financial support of Labex CheMISyst and UM.

- **REFERENCES**

1. M. Stemmelen, C. Travelet, V. Lapinte, R. Borsali and J. J. Robin, *Polymer Chemistry*, 2013, **4**, 1445-1458.
2. M. Stemmelen, V. Lapinte, J. P. Habas and J. J. Robin, *European Polymer Journal*, 2015, **68**, 536-545.
3. P. D. Pham, V. Lapinte, Y. Raoul and J. J. Robin, *Journal of Polymer Science Part a-Polymer Chemistry*, 2014, **52**, 1597-1606.
4. A. M. Diez-Pascual and A. L. Diez-Vicente, *Biomacromolecules*, 2015, **16**, 2631-2644.
5. G. Lligadas, J. C. Ronda, M. Galia and V. Cadiz, *Biomacromolecules*, 2010, **11**, 2825-2835.
6. C. Q. Zhang, Y. Xia, R. Q. Chen, S. Huh, P. A. Johnston and M. R. Kessler, *Green Chemistry*, 2013, **15**, 1477-1484.
7. N. B. Tran, J. Vialle and Q. T. Pham, *Polymer*, 1997, **38**, 2467-2473.
8. B. S. Madhukar, D. G. B. Gowda, V. Annadurai, R. Somashekar and Siddaramaiah, *Advances in Polymer Technology*, 2016, **35**, 21526-21536.
9. S. Allauddin, R. Narayan and K. Raju, *Acs Sustainable Chemistry & Engineering*, 2013, **1**, 910-918.
10. Y. Mulazim, E. Cakmakci and M. V. Kahraman, *Progress in Organic Coatings*, 2011, **72**, 394-401.
11. M. A. de Luca, M. Martinelli, M. M. Jacobi, P. L. Becker and M. F. Ferrao, *Journal of the American Oil Chemists Society*, 2006, **83**, 147-151.
12. R. Guo, X. Du, R. Zhang, L. Deng, A. Dong and J. Zhang, *European Journal of Pharmaceutics and Biopharmaceutics*, 2011, **79**, 574-583.
13. P. T. Knight, K. M. Lee, T. Chung and P. T. Mather, *Macromolecules*, 2009, **42**, 6596-6605.
14. R. J. Zdrahala and I. J. Zdrahala, *Journal of Biomaterials Applications*, 1999, **14**, 67-90.
15. B. D. Ulery, L. S. Nair and C. T. Laurencin, *Journal of Polymer Science Part B-Polymer Physics*, 2011, **49**, 832-864.

16. T. Tsujimoto, H. Uyama and S. Kobayashi, *Macromolecular Rapid Communications*, 2003, **24**, 711-714.
17. H. Schmidt, *Journal of Non-Crystalline Solids*, 1989, **112**, 419-423.
18. N. Devia, J. A. Manson, L. H. Sperling and A. Conde, *Macromolecules*, 1979, **12**, 360-369.
19. M. A. de Luca, M. Martinelli and C. C. T. Barbieri, *Progress in Organic Coatings*, 2009, **65**, 375-380.
20. G. Lligadas, J. C. Ronda, M. Galia and V. Cadiz, *Biomacromolecules*, 2006, **7**, 3521-3526.
21. S. J. Tuman and M. D. Soucek, *Journal of Coatings Technology*, 1996, **68**, 73-81.
22. R. A. Sailer, J. R. Wegner, G. J. Hurtt, J. E. Janson and M. D. Soucek, *Progress in Organic Coatings*, 1998, **33**, 117-125.
23. G. H. Teng, J. R. Wegner, G. J. Hurtt and M. D. Soucek, *Progress in Organic Coatings*, 2001, **42**, 29-37.
24. A. E. Atabani, A. S. Silitonga, H. C. Ong, T. M. I. Mahlia, H. H. Masjuki, I. A. Badruddin and H. Fayaz, *Renewable & Sustainable Energy Reviews*, 2013, **18**, 211-245.
25. A. S. Trevino and D. L. Trumbo, *Progress in Organic Coatings*, 2002, **44**, 49-54.
26. C. Q. Zhang, S. A. Madbouly and M. R. Kessler, *Acs Applied Materials & Interfaces*, 2015, **7**, 1226-1233.
27. R. Irwin, *Toxicity studies of castor oil in F344/N rats and B6C3F1 mice-Technical Report*, 1992.
28. F. C. Naughton, *Journal of the American Oil Chemists Society*, 1974, **51**, 65-71.
29. R. Dave, Overview of pharmaceutical excipients used in tablets and capsules, http://drugtopics.modernmedicine.com/drugtopics/Top+News/Overview-of-pharmaceutical-excipients-used-in-tabl/ArticleStandard/Article/detail/561047, Accessed 13/03/2017.
30. K. Yamamoto and N. Fujiwara, *Bioscience Biotechnology and Biochemistry*, 1995, **59**, 1262-1266.
31. K. C. Frisch and D. Klempner, eds., *Advances in Urethane Science and Technology*, 1996.
32. G. Trovati, E. Ap Sanches, S. C. Neto, Y. P. Mascarenhas and G. O. Chierice, *Journal of Applied Polymer Science*, 2010, **115**, 263-268.
33. D. M. Bechi, M. A. de Luca, M. Martinelli and S. Mitidieri, *Progress in Organic Coatings*, 2013, **76**, 736-742.
34. Y. Mulazim, E. Cakmakci and M. V. Kahraman, *Journal of Vinyl & Additive Technology*, 2013, **19**, 31-38.
35. S. Miao, P. Wang, Z. Su and S. Zhang, *Acta Biomaterialia*, 2014, **10**, 1692-1704.
36. G. Gallon, V. Lapinte, J.-J. Robin, J. Chopineau, J.-M. Devoisselle and A. Aubert-Pouëssel, *Sustainable Chemistry and Engineering B*, 2013, **In press**.
37. D. H. Kaelble, *Journal of Adhesion*, 1970, **2**, 66-81.
38. D. Owens and R. Wendt, *J. Appl. Polym. Sci* 1969, **13**, 1741-1747.
39. K. M. S. Meera, R. M. Sankar, S. N. Jaisankar and A. B. Mandal, *Journal of Physical Chemistry B*, 2013, **117**, 2682-2694.
40. D. Fischer, D. Pospiech, U. Scheler, R. Navarro, M. Messori and P. Fabbri, *Macromolecular Symposia*, 2008, **265**, 134-143.
41. A. Ali, K. Yusoh and S. F. Hasany, *Journal of Nanomaterials*, 2014, 1-10.
42. N. de la Rosa-Fox, V. Morales-Florez, J. A. Toledo-Fernandez, M. Pinero, R. Mendoza-Serna and L. Esquivias, *Journal of the European Ceramic Society*, 2007, **27**, 3311-3316.

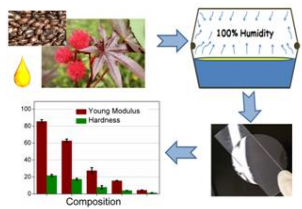

**TOC:** Castor oil-based hybrid film formation via sol-gel chemistry with tunable mechanical properties.